\documentclass[aps,prx,showpacs,superscriptaddress, twocolumn]{revtex4-2}
\usepackage{graphicx}
\usepackage{dcolumn}
\usepackage{bm}
\usepackage{amsmath}
\usepackage{amsfonts}
\usepackage{amssymb}
\usepackage{hyperref}
\usepackage{mathtools}
\usepackage{braket}
\usepackage{comment}
\usepackage{multirow}
\usepackage{here} 
\usepackage{bm}
\usepackage{color}
\usepackage{xcolor}
\usepackage{siunitx}
\usepackage{amsthm}
\theoremstyle{plain}

\begin{document}

\preprint{APS/123-QED}%

\title{General treatment of Gaussian trusted noise \\
in continuous variable quantum key distribution}

\author{Shinichiro Yamano}%
\email{yamano@qi.t.u-tokyo.ac.jp}
\affiliation{
Department of Applied Physics, Graduate School of Engineering,The University of Tokyo, 7-3-1 Hongo Bunkyo-ku, Tokyo 113-8656, Japan}

\author{Takaya Matsuura}
\affiliation{
Centre for Quantum Computation $\&$ Communication Technology, School of Science, RMIT University, Melbourne VIC 3000, Australia}

\author{\\Yui Kuramochi}
\affiliation{
Department of Physics, Kyushu University, 744 Motooka, Nishi-ku, Fukuoka, Japan}

\author{Toshihiko Sasaki}
\affiliation{
Department of Applied Physics, Graduate School of Engineering,The University of Tokyo, 7-3-1 Hongo Bunkyo-ku, Tokyo 113-8656, Japan}
\affiliation{
Photon Science Center, Graduate School of Engineering,The University of Tokyo, 7-3-1 Hongo, Bunkyo-ku, Tokyo 113-8656, Japan}

\author{Masato Koashi}
\affiliation{
Department of Applied Physics, Graduate School of Engineering,The University of Tokyo, 7-3-1 Hongo Bunkyo-ku, Tokyo 113-8656, Japan}
\affiliation{
Photon Science Center, Graduate School of Engineering,The University of Tokyo, 7-3-1 Hongo, Bunkyo-ku, Tokyo 113-8656, Japan}

\begin{abstract}%
Continuous Variable (CV) quantum key distribution (QKD) is a promising candidate for practical implementations due to its compatibility with the existing communication technology.
A trusted device scenario assuming that an adversary has no access to imperfections such as electronic noises in the
detector is expected to provide significant improvement in the key rate, but such an endeavor so far was made separately for specific protocols and for specific proof techniques.
Here, we develop a simple and general treatment that can incorporate the effects of Gaussian trusted noises for any protocol that uses homodyne/heterodyne measurements.
In our method, a rescaling of the outcome of a noisy homodyne/heterodyne detector renders it equivalent to the outcome of a noiseless detector with a tiny additional loss, thanks to a noise-loss equivalence well-known in quantum optics. Since this method is independent of protocols and security proofs, 
it is applicable to  Gaussian-modulation and discrete-modulation protocols, to the finite-size regime,  and to any proof techniques developed so far and yet to be discovered
as well. 
\end{abstract}
\maketitle

\section{Introduction}

\begin{figure*}[t]
  \centering
  \begin{minipage}{1\columnwidth}
    \centering
    \includegraphics[width=\columnwidth]{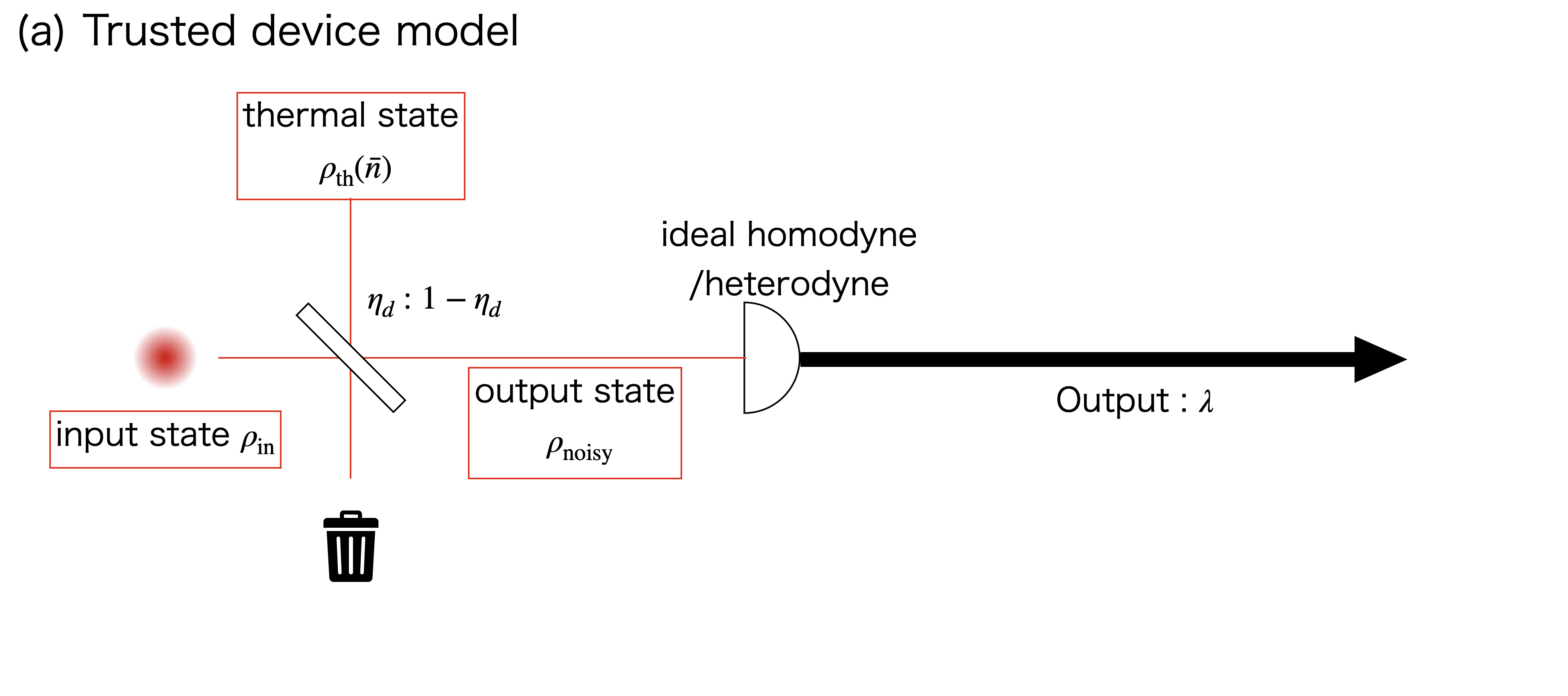}\\
    \includegraphics[width=\columnwidth]{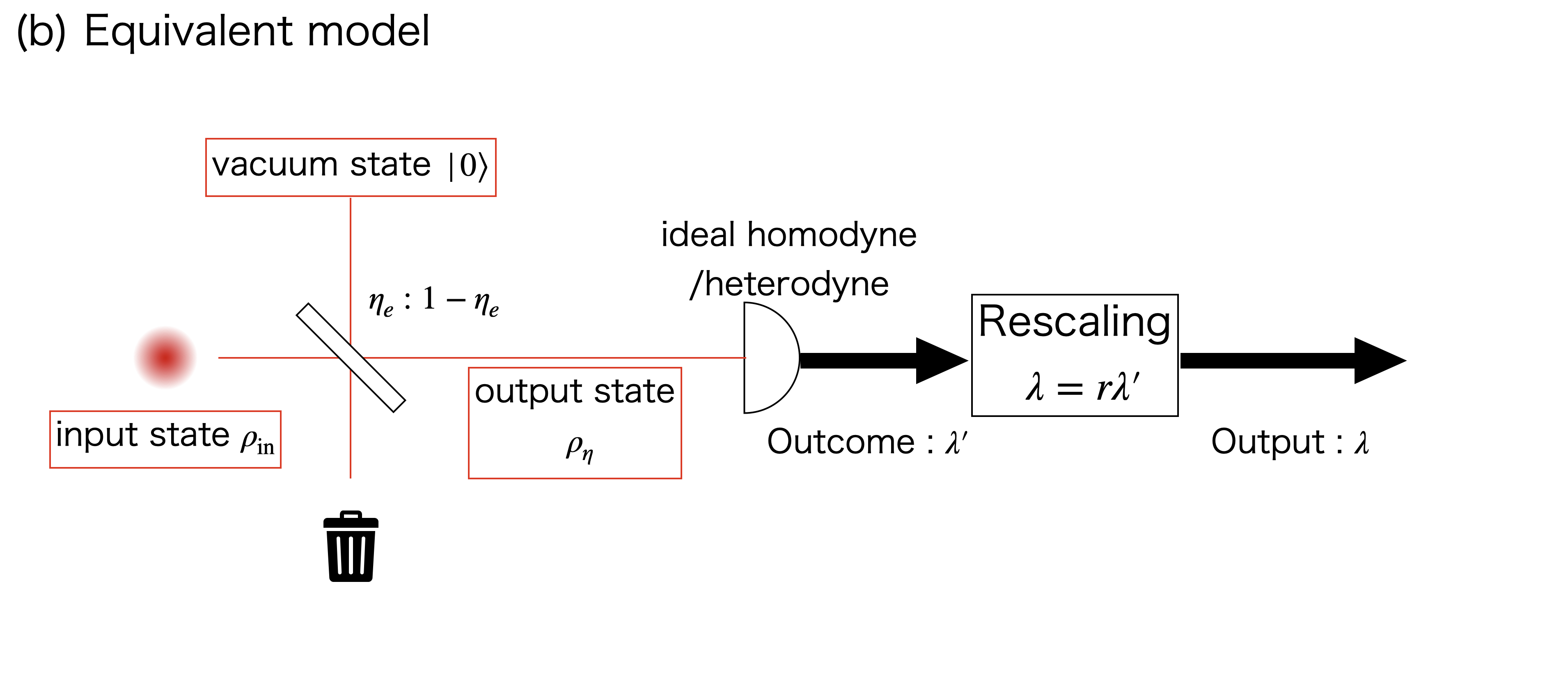}
    \caption{(a) Trusted device model of a noisy measurement in CV QKD. (b) Measurement model equivalent to (a).} 
    \label{fig1}
  \end{minipage}
  \hfill
  \centering
  \begin{minipage}{1\columnwidth}
    \centering
    \includegraphics[width=\columnwidth]{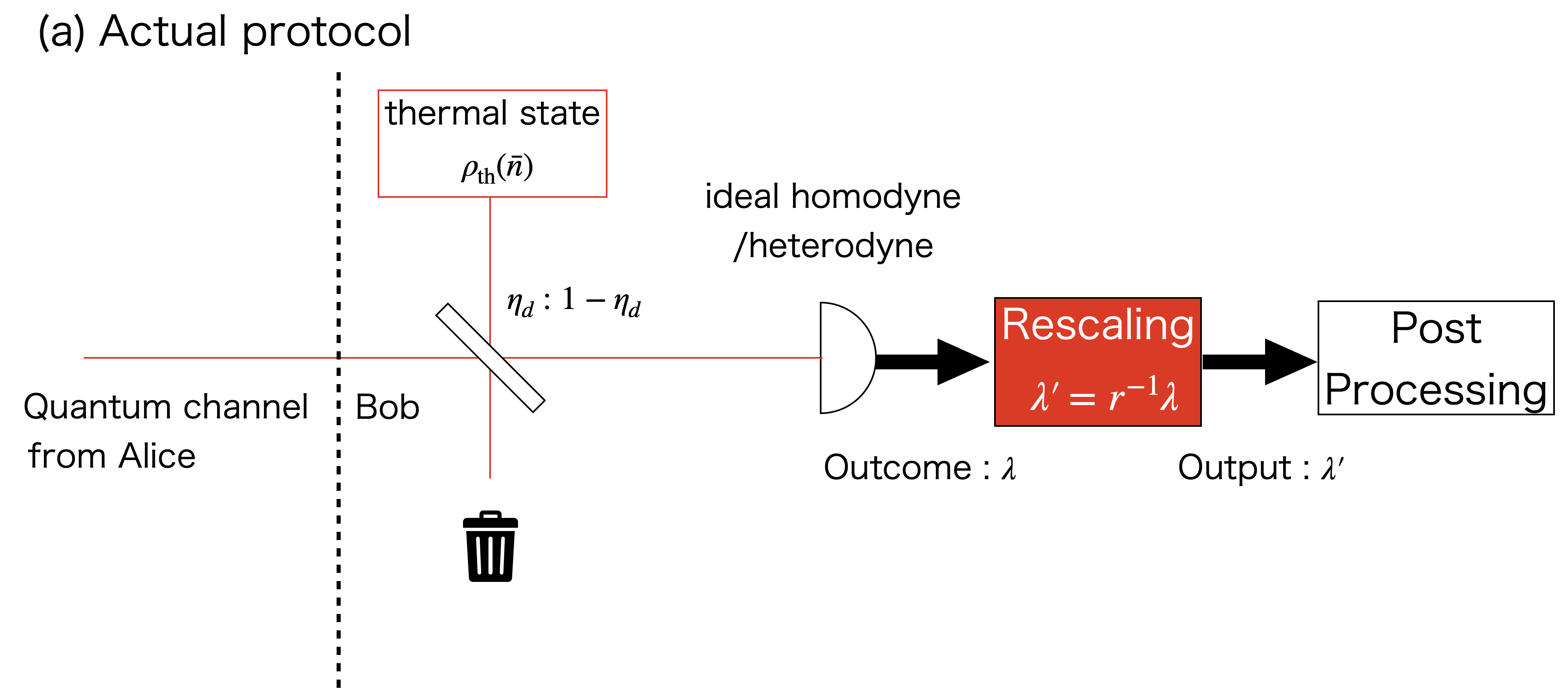}\\
    \includegraphics[width=\columnwidth]{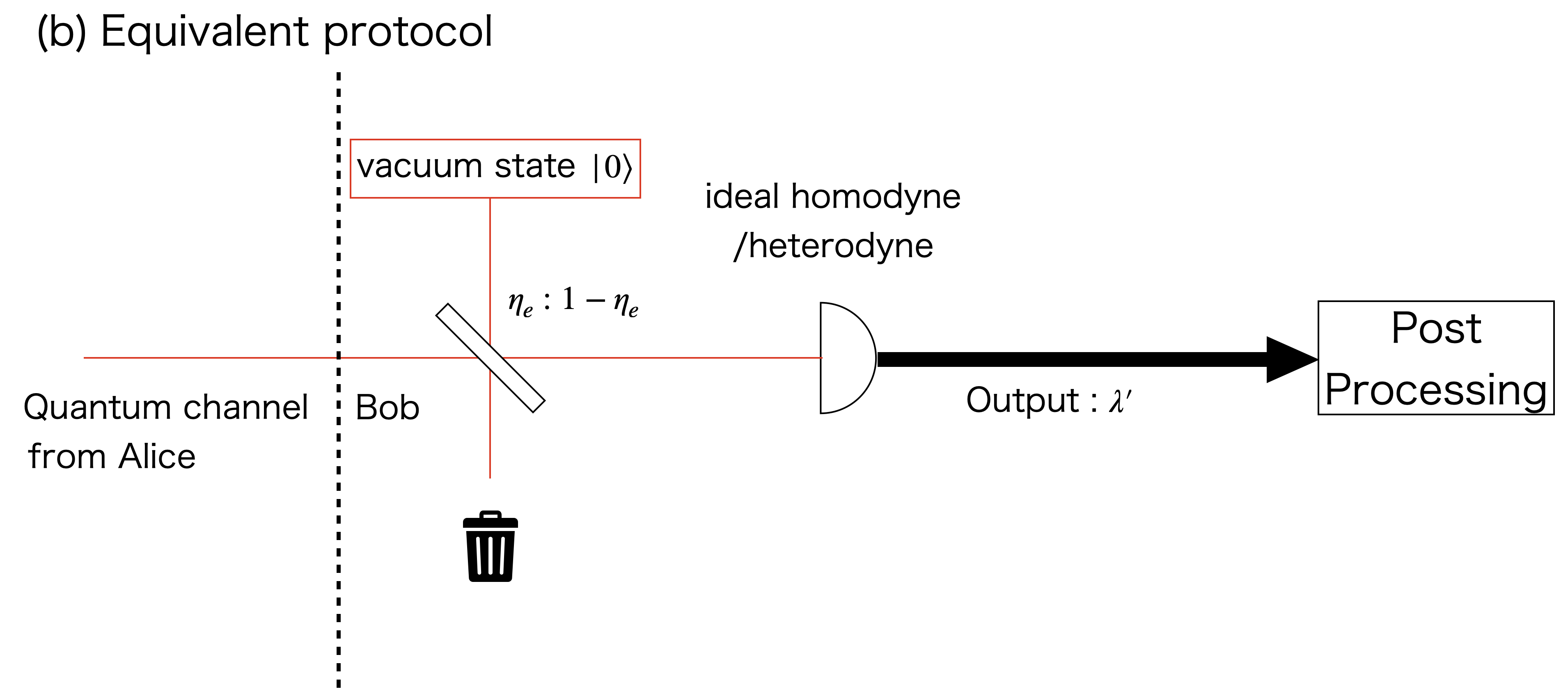}
    \caption{(a) Actual protocol with a noisy measurement followed by rescaling. (b) Equivalent protocol with a noiseless measurement.}
    \label{fig2}
  \end{minipage}
\end{figure*}

Quantum key distribution (QKD) is the technology that enables information-theoretically secure communication between two separate parties.
QKD is classified into discrete-variable (DV) QKD and continuous-variable (CV) QKD. 
It is customary to make the classification according to detection methods in the receiver, regardless of encoding methods in the transmitter.
DV-QKD protocols use photon detectors to read out the encoded information.
The early proposals \cite{BENNETT1984,Bennett1992} of QKD belong to this type, and much
knowledge about finite-key analysis and how to handle imperfections of actual devices has been accumulated.
On the other hand, CV-QKD protocols 
use homodyne or heterodyne detection \cite{Ralph1999,Grosshans2002},
which is highly compatible with the coherent optical communication technology currently widespread in the industry \cite{eriksson2020wavelength, huang2015continuous, kumar2015coexistence, huang2016field, karinou2017experimental, karinou2018toward, eriksson2018coexistence, eriksson2019wavelength}.

In actual experiments, the state sent by the sender Alice is not directly transmitted to the receiver Bob, but  the state is affected by noises and losses in the quantum channel and their devices. 
As for the noises and losses that occurred in the quantum channel, Alice and Bob cannot distinguish whether they are caused by the eavesdropper Eve's attack. 
One must then assume that they are all caused by Eve 
in calculating an appropriate amount of privacy amplification.
On the other hand, since the detector is physically protected in Bob's lab, it is often reasonable to assume that Eve has no access to 
imperfections in the detector \cite{Pirandola2021}.
The security under this relaxed assumption has been considered in the past and is called a trusted device scenario (see review \cite{Usenko2016}). 
Especially for homodyne/heterodyne measurements used in CV-QKD, where the classical noise of the electronic circuit after the detection is often larger than the excess noise of the channel, the trusted device assumption can significantly improve the key rate of a QKD protocol \cite{ Lodewyck2007, grosshans2003quantum, Fossier2009_2, jouguet2013experimental,huang2016long, ren2021demonstration}. 

From a theoretical point of view, 
it is crucial to find quantitative relations between the parameters describing the amount of trusted noises and 
the reduction in the amount of privacy amplification by assessing how the trusted device scenario limits the
amount of information leaked to Eve. 
So far, such an endeavor was made separately for specific protocols and for  specific proof techniques.
In Gaussian modulation schemes, since it was proved that the optimal attack for Eve is the entangling cloner attack, it is relatively easy to evaluate the effect of 
Gaussian trusted noises by computing the covariance matrix of a quantum state before it incurred the trusted noises \cite{huang2013quantum,ma2014enhancement,PhysRevA.86.032309,Laudenbach2018,Laudenbach2019, ren2019reference, Pirandola2021,shao2021phase,kunz2015robust}.
As for discrete modulation protocols,
Namiki \textit{et al.}\cite{Namiki2018} adopted an
assumption that limits Eve's attacks to the entangling cloner attacks
and determined the amount of privacy amplification for Gaussian noises. 
Lin and L\"{u}tkenhaus \cite{Lin2020} relaxed the assumption to collective attacks in the case of a specific proof technique based on 
numerical optimization.
Currently, there are many different approaches \cite{Leverrier2017, m, Entacc} to the finite-size security against general attacks on CV-QKD, but the trusted device scenario for these approaches has not been explicitly given. Mere extension of existing methods would require a separate argument for each of the finite-size approaches.

In this study, we develop a simple and general treatment that can incorporate the effects of Gaussian trusted noises for any protocol that applies homodyne/heterodyne measurements.
Specifically, we introduce a rescaling of the outcome of a noisy homodyne/heterodyne detector which renders it equivalent to a
noiseless detector with a tiny additional loss, thanks to the noise-loss equivalence well-known in quantum optics.
Since this method is independent of protocols and security proofs, it is applicable 
to Gaussian-modulation and discrete modulation protocols, to asymptotic and finite-sized regimes, and to any proof techniques developed
so far and yet to be discovered as well.


This paper is organized as follows. In Sec.~\ref{sec:method}, 
we introduce 
an equivalent model of the noisy measurement
in the trusted device scenario and explain a method for treating trusted Gaussian noises for homodyne/heterodyne detection.
In Sec.~\ref{sec:example}, 
we specifically apply our method to the binary modulation protocol and show numerical simulations of the key rates.
We conclude this paper in Sec.~\ref{sec:conclusion}.

\section{\label{sec:method}METHODS}

The main imperfection in the homodyne/heterodyne measurement device is its non-unit quantum efficiency and noises on the electronic circuit.
As in the previous analyses on the trusted noises, we assume that the noises are Gaussian, and that the imperfect homodyne/heterodyne measurement is equivalently represented by a quantum optical model shown
in Fig.\;\ref{fig1}\;(a).
In this model, the input optical pulse is mixed with an ancillary pulse $R$ that is  in a thermal state with an average photon number $\bar{n}$ by a beam splitter with a transmittance $\eta_d$, and then is fed to an ideal homodyne/heterodyne detector to produce an outcome $\lambda\in \Omega$ ($\Omega=\mathbb{R}$ for homodyne and $\Omega = \mathbb{C}$ for heterodyne).
In the trusted device scenario, we assume that Eve has no access to the device and hence the thermal ancillary pulse
and Eve's system are decoupled.

Let $\rho$ be the state of the input pulse and $\rho_{\rm noisy}$ be the state of the pulse just before the 
ideal homodyne/heterodyne detector.
The completely positive and trace preserving (CPTP) map $\mathcal{E}$ for the mixing of the thermal light
is represented by
\begin{align}
 \rho_{\rm noisy} =\mathcal{E}(\rho) = {\rm Tr}_R \left[ S(\eta_d) \rho \otimes \rho_{\rm th}(\bar{n}) S^\dagger(\eta_d)\right] ,
\end{align}
where $\rho_{\rm th}(\bar{n})$ is a thermal state with the mean photon number $\bar{n}$ and $S(\eta)$ is the unitary operator for the beamsplitter with transmittance $\eta$ .
Let us denote the positive-operator-valued measure (POVM) for the ideal homodyne/heterodyne detector as $\Pi(\Delta)(\Delta \subset \Omega)$.
The POVM $\Pi_{\rm noisy}(\Delta)$ 
of the noisy homodyne/heterodyne measurement modeled in Fig. \ref{fig1} (a) is then given by 
\begin{align}
\Pi_{\rm noisy}(\Delta) = \mathcal{E}^\dagger(\Pi(\Delta)).
\end{align}
where $\mathcal{E}^\dagger$ is the adjoint map of $\mathcal{E}$.

Next, we consider another model shown in Fig.\;\ref{fig1} (b), which will be shown to be equivalent to the one in Fig.\;\ref{fig1} (a).
Here, the input light pulse is mixed with the vacuum state $\ket{0}$ using a beamsplitter with a transmittance $\eta_e$
and is then fed to the same  homodyne/heterodye detector to produce an intermediate outcome $\lambda' \in \Omega$. 
It is then rescaled by a factor $r \ge 1$ to produce a final outcome $\lambda=r \lambda'$.
Let  $\rho_{\eta_e}$ be the state of the pulse just before the ideal homodyne/heterodyne detector. The CPTP map for converting $\rho$ to $\rho_{\eta_e}$ is then given by
\begin{align}
   \rho_{\eta_e}  =  \mathcal{E}_{\rm eq}(\rho) = {\rm Tr}_R \left[ S(\eta_e)  \rho \otimes \ket{0}\bra{0}_R S^\dagger(\eta_e) \right].
\end{align}
The POVM $ \Pi_{\rm eq}(\Delta) $ of the measurement modeled in Fig.\;\ref{fig1} (b) is then given by 
\begin{align}
 \Pi_{\rm eq}(\Delta) = \mathcal{E}_{\rm eq}^\dagger(\Pi(r^{-1}(\Delta))),
\end{align}
where $r^{-1}(\Delta)=
\{\lambda' | r \lambda'\in \Delta\}$.
The measurement model in Fig.\;\ref{fig1} (b) can be made equivalent to the one in  Fig.\;\ref{fig1} (a) by appropriately 
choosing $\eta_e$ and $r$. 
We will prove in Appendix \ref{homo}, \ref{hetero} that $\Pi_{\rm noisy}(\Delta) = \Pi_{\rm eq}(\Delta) $ holds when
\begin{align}
r^2 &= 1+2\bar{n}(1-\eta_d) \label{homo_v}\\
\eta_e &=r^{-2}  \eta_d  . \label{homo_res}
\end{align}
for a homodyne measurement, and when
\begin{align}
 r^2 &= 1+\bar{n}(1-\eta_d) \label{hetero_v}\\
\eta_e &= r^{-2}  \eta_d .  \label{hetero_res}
\end{align}
for a heterodyne measurement.

Our general method for trusted Gaussian noises is stated as follows. Whenever a QKD protocol dictates to use
an outcome $\lambda$ of a homodyne/heterodyne detector, we modify it to use its rescaled value $\lambda'= r^{-1} \lambda$
instead of $\lambda$. If the parameter $r$ is chosen according to the parameters $(\eta_d, \bar{n})$ of the trusted noises of the detector, this modified protocol in Fig.\;\ref{fig2}\;(a) is equivalent to the one in Fig.\;\ref{fig2}\;(b). This enables us to use any security proof assuming noiseless detector with a slightly increased loss $\eta_e=r^{-2}\eta_d$
to determine a sufficient amount of privacy amplification. 

When a CV QKD protocol uses two or more homodyne/heterodyne detectors, the above method may not work as it is, because security proofs often assume that different detectors have the same quantum efficiency. For example,
a security proof is first constructed assuming ideal homodyne/heterodyne detectors, and then is claimed to be
applicable to the case with inefficient (but noiseless) detectors with the same quantum efficiency, on the basis that
the common loss can be ascribed to Eve's attack.

Fortunately, there are several remedies at the implementation level. Since the reduced efficiency $\eta_e$ is
a function of original loss $\eta_d$ and noise $\bar{n}$, one may introduce an additional loss or noise to
a detector to decrease its reduced efficiency $\eta_e$. For an additional loss, one may directly
insert a lossy component or may intentionally worsen the spatial mode matching between the signal and LO.
For an additional noise, one may use random-number generator to add a Gaussian noise to the outcome,
or may just reduce the LO power if the original noise is in the photodetector and the subsequent circuits.
Given the worst value $\eta_e^{\rm (min)}$ among the detectors,
one can thus make all the detectors have the reduced efficiency $\eta_e^{\rm (min)}$ in this way.

\begin{figure}[h]
  \centering
    \includegraphics[width=0.9\columnwidth]{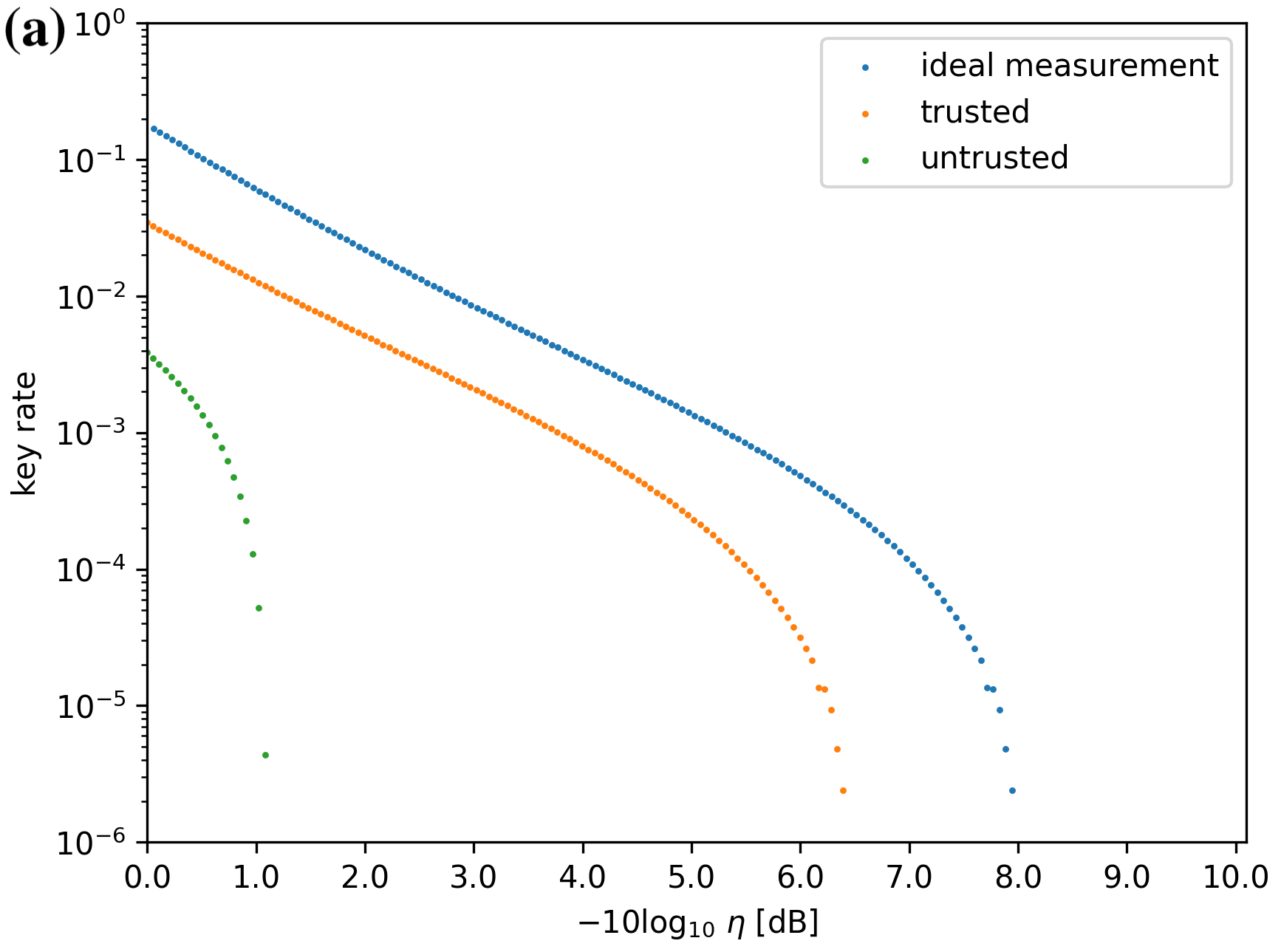}
    \includegraphics[width=0.9\columnwidth]{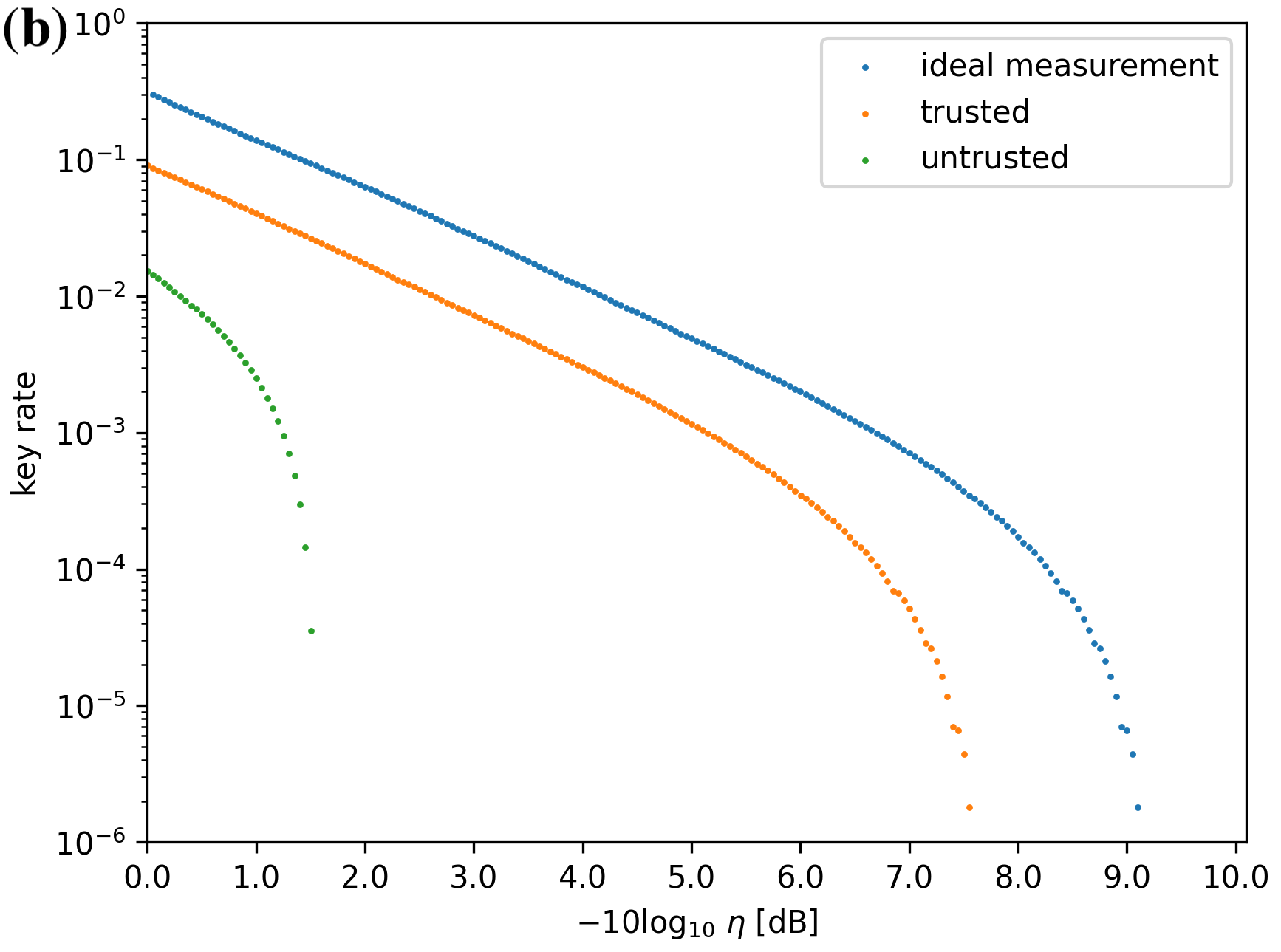 }
  \caption{The key rate comparison for the the cases where the detector is ideal, has trusted noises, and has untrusted noises. (a) Protocol with heterodyne measurement. (b) Protocol with homodyne and heterodyne measurements. }    
  \label{3}
\end{figure}

\section{\label{sec:example}Example}

As an example, we apply this method to two variants of reverse reconciliation protocols with binary modulation \cite{matsuura2023}.
Two variants of protocols were analyzed in Ref.\;\cite{matsuura2023}, an all-heterodyne protocol and a hybrid protocol using both homodyne and heterodyne measurements. 
For both protocols, the security proof covers the finite-key-size regime against general coherent attacks.
For this protocol, the conventional treatment of the trusted noise\cite{huang2013quantum,ma2014enhancement,PhysRevA.86.032309,Laudenbach2018,Laudenbach2019, ren2019reference, Pirandola2021,shao2021phase,kunz2015robust} does not work since it does not use the property of the covariance matrix.

For the simulation of the key rate, we set the security parameter $\epsilon_{\rm sec} = 2^{-50}$ and the communication channel as follows.
We assume the communication channel to be a phase-invariant Gaussian channel with the transmittance $\eta$ and excess noise $\xi$. 
We assume that the noise is ascribed to a fixed noise source at the input with
$\xi_0 = 10^{-3}$, which means the excess noise is given by $\xi = \eta\xi_0$. This quantum channel can be modeled by random displacement 
followed by a pure loss channel.
Under this model, a coherent state $\ket{\beta}$, after transmitted through this quantum channel, becomes
\begin{align}
 \rho_{\rm in} = \frac{1}{\eta\xi_0}\int_{\mathbb{C}} \frac{2}{\pi}   \exp\left( -\frac{2|\gamma|^2}{\eta\xi_0}\right) \ket{\sqrt{\eta}\beta+\gamma} \bra{\sqrt{\eta}\beta+\gamma}d^2\gamma.
\end{align}

For the all-heterodyne protocol, we assume that the heterodyne detector has a
quantum efficiency $\eta_d = 0.7=1.5 \si{\decibel}$ and noise $2\bar{n}(1-\eta_d) = 10^{-3}$ and simulated the finite-size key rates with $10^{12}$ transmitted pulses.
In the trusted noise scenario, our method gives $r^2=1+0.5\times 10^{-3}$ and the key rate is the same as
the case with a Gaussian channel with transmittance $\eta \eta_e=\eta \eta_d r^{-2}$ and excess noise $\eta\eta_e\xi_0$ followed by an ideal detector, 
which can be calculated using the formula in Ref.\;\cite{matsuura2023} and is shown as the curve labeled as ``trusted'' 
in Fig.\;\ref{3} (a). If we do not use the trusted scenario and ascribe all the imperfection to the 
quantum channel, it becomes a Gaussian channel with transmittance $\eta \eta_d$ and excess noise $\eta\eta_d\xi_0 + 10^{-3}$.
This rate is shown as the curve labeled as ``untrusted'' 
in Fig.\;\ref{3} (a). We also show the key rate when the ideal detector ($\eta_d=\bar{n}=0$) is used.
We see that the trusted-device scenario significantly improves the key 
rate compared to the untrusted one, 
to the extent that the effect of the noises is almost eliminated.


For the hybrid protocol, we consider the case where both of the detectors have the same 
quantum efficiency $\eta_d=0.7=1.5 \si{\decibel}$ and the same noise parameter $2\bar{n}(1-\eta_d)=10^{-3}$.
Then the rate with the untrusted case is again given by considering a quantum channel made up of
 a Gaussian channel with transmittance $\eta \eta_d$ and excess noise $\eta\eta_d\xi_0 +10^{-3}$.
For the trusted scenario, the reduced efficiency differs for the two detectors:
$\eta_e=\eta_d/(1+10^{-3})$ for the homodyne detector and $\eta_e=\eta_d/(1+0.5\times 10^{-3})$ 
for the heterodyne detector. Hence our method is applicable after an additional loss or noises are
introduced such that its reduced efficiency becomes $\eta_e^{\rm (min)}=\eta_d/(1+10^{-3})$.
The key rate is then given as the case with a Gaussian channel with transmittance $\eta \eta_e^{\rm (min)}$ and excess noise $\eta \eta_e^{\rm (min)}\xi_0$ followed by an ideal detector. The result is shown in Fig.\;\ref{3} (b), which shows almost the same tendency as Fig.\;\ref{3} (a).

\section{\label{sec:conclusion}discussion and conclusion}
As shown in Fig.\;\ref{3},
in both the homodyne and hybrid protocols, the difference between an ideal detector and a detector with trusted noise is essentially only in the detection efficiency.
This behavior can be understood in the general terms as follows. The noise-loss equivalence 
in a homodyne/heterodyne measurement trades an excess noise $\xi$ with an additional loss of $O(\xi)$.
Under this equivalence, the performance of the CV-QKD is affected by the two types of imperfections 
in an extremely asymmetric way, namely, it is vulnerable against noise but robust against loss.
This is the reason why the trusted noise scenario remarkably improves the key rate of CV-QKD.


It is worth mentioning that assuming the device model of Fig.\;\ref{fig1} (a) with parameters $\eta_d$ and $\bar{n}$ 
does not necessarily require the characterization of the actual detector for the two parameters.
It is often the case that a QKD protocol (including the computation of the amount of privacy amplification)
does not use the value of $\eta_d$. In such a case, the only value required in constructing the Actual protocol 
in Fig.\;\ref{fig2} (a) in our method is $\bar{n}(1-\eta_d)$.
Characterization of this value is achieved by inputting a vacuum state into the device and measuring the variance of its quadrature.

There is also a minor but frequent issue in the model of Fig.\;\ref{fig1} (a). 
In QKD, it is customary to model the actual detector by a pure-loss channel followed by a detector with $\eta_d=1$ while the channel is assumed to be in Eve's control. It is preferred (i) when an available security proof assumes a detector with perfect efficiency, or (ii) when the apparent loss in the detector cannot be trusted and 
Eve may suppress it.
In either case, there are cases when we can reasonably trust the noises in the detector, such as the electrical noises. A problem in applying our method of trusted scenario is that the model of Fig.\;\ref{fig1} (a) with $\eta_d=1$ cannot represent nonzero noises. 
The remedy here is to represent a noisy detector with $\eta_d=1$ by a limit of a sequence of models of Fig.\;\ref{fig1} (a)
with a fixed value of $\bar{n}(1-\eta_d)$ and $\eta_d \to 1$. 
This allows the computation of $r$ and $\eta_e$ to apply our method.


In conclusion, we propose a simple method to treat Gaussian noise in the trusted device scenario. 
By introducing a slight modification in the actual protocol 
such that one uses rescaled values of quadrature outcomes of the noisy detectors, 
we can benefit from a device-level equivalence to noiseless detectors with a tiny
additional loss.
CV-QKD protocols tend to be vulnerable to noise, and it has been suggested that 
the trusted device scenario for the receiver markedly improves the key rates.
Our contribution is to widen its applicability
considerably without regard to protocols and proof
techniques. We numerically confirmed for specific protocols with 
finite-size security proofs that the trusted device scenario 
almost entirely mitigates the effect of the detector noise.
Since our method has a simple effect of trading the Gaussian 
detector noise with a tiny loss, we expect a similar benefit 
from any CV-QKD protocol with the trusted device scenario. 

\section*{Acknowledgement}
This work was supported by 
Cross-ministerial Strategic Innovation Promotion Program (SIP) (Council for Science, Technology and Innovation (CSTI)); 
the Ministry of Internal Affairs and Communications (MIC) under the initiative Research and Development for Construction of a Global Quantum Cryptography Network (grant number JPMI00316);
JSPS  Grants-in-Aid  for  Scientific  Research No.~JP22K13977;
JST Moonshot R\&D, Grant Number JPMJMS2061;
JSPS Overseas Research Fellowships.

\appendix

\section{homodyne measurement}\label{homo}

We introduce the annihilation operator $\hat{a}$ and creation operator $\hat{a}^\dagger$ of a single-mode state with the usual commutation relation $[\hat{a},\hat{a}^\dagger] = 1$, and define the quadrature operator:
\begin{align}
\hat{x} &= \frac{1}{2}(\hat{a}+\hat{a}^\dagger) \\
\hat{p} &= \frac{i}{2}(\hat{a}-\hat{a}^\dagger).
\end{align}
The coherent  state $\ket{\alpha}$ is written as
\begin{align}
\ket{\alpha} =  e^{-\frac{|\alpha|^2}{2}}\sum_{n=0}^{\infty} \frac{\alpha^n}{\sqrt{n!}}\ket{n},
\end{align}
and the wave function $\braket{x|\alpha}$ is
\begin{align}
 \braket{x|\alpha} = \left( \frac{2}{\pi} \right)^{\frac{1}{4}} \exp[-(x-\alpha_R)^2 + 2i \alpha_I x-i\alpha_R\alpha_I],
\end{align}
where $\alpha_R = {\rm Re}(\alpha),\alpha_I = {\rm Im}(\alpha)$.
The density operator of the thermal state with mean photon number $\bar{n}$ is described as
\begin{align}
 \rho_{\rm th}(\bar{n}) &=  \sum_{n=0}^\infty \frac{\bar{n}^n}{(1+\bar{n})^{n+1}} \ket{n}\bra{n}\\
&=  \frac{1}{\pi \bar{n}}\int_\mathbb{C}\exp(-\frac{|\beta|^2}{\bar{n}})\ket{\beta}\bra{\beta}   d^2\beta.
\end{align}
The condition for two POVMs to be equal is that their probability distributions are equal for any coherent states \cite{PhysRev.138.B274}.
Then, we consider the coherent state $\rho_{\rm in} = \ket{\alpha}\bra{\alpha}$ as input.
In the trusted device model, the state to which Bob performs the ideal measurement is calculated as
\begin{align}
 \mathcal{E}( \rho_{\rm in} )&= \rho_{\rm noisy}\\ 
&=  {\rm Tr}_R  \left[ S(\eta_d)  \ket{\alpha}\bra{\alpha} \otimes  \rho_{\rm th}(\bar{n}) S^\dagger(\eta_d) \right] \\
&= \frac{1}{\pi} \int  \frac{1}{\bar{n}} \exp\left(- \frac{|\beta|^2}{\bar{n}} \right) \nonumber \\
& \ket{\sqrt{\eta_d}\alpha+ \sqrt{1-\eta_d}\beta }\bra{\sqrt{\eta_d}\alpha+ \sqrt{1-\eta_d}\beta }   d^2\beta ,
\end{align}
and  in the equivalent model, Bob measures the state  $\mathcal{E}_{\rm eq}(\rho_{\rm in})=\rho_{\eta_e} = \ket{\sqrt{\eta_e}\alpha}\bra{\sqrt{\eta_e}\alpha}$.

The homodyne measurement outputs a real number $x\in \mathbb{R}$.
The probability density of an outcome $x$ of an ideal homodyne measurement with an input state $\rho$ is given by $\braket{x|\rho|x}$. 
The probability distribution obtained by the trusted device model ${\rm Pr}\left( x\in \Delta \right)$ is written as
\begin{align}
&{\rm Tr}(\rho_{\rm in} \Pi_{\rm noisy}(\Delta))  \\
=& {\rm Tr}( \mathcal{E}(\rho_{\rm in}) \Pi(\Delta)) \\
=&\int_{x\in\Delta}  \bra{x} \mathcal{E}(\rho_{\rm in}) \ket{x} dx\\
=&\int_{x\in\Delta}  \frac{1}{\sqrt{2\pi\sigma^2_{\rm hom}}}\exp (-\frac{(x-\sqrt{\eta_d}x_0)^2}{2 \sigma^2_{\rm hom} })dx,
\end{align}
where we defined $x_0 = {\rm Re}(\alpha)$ and
\begin{align}
 \sigma^2_{\rm hom} & = \frac{1+2 \bar{n} (1-\eta_d)}{4}.
\end{align}
Then, in the equivalent model, the probability distribution is given by
\begin{align}
&{\rm Tr}(\rho_{\rm in} \Pi_{\rm eq}(\Delta))  \\
=& {\rm Tr}( \mathcal{E}_{\rm eq}(\rho_{\rm in}) \Pi(r^{-1}(\Delta))) \\
=&\int_{x'\in r^{-1} (\Delta)}  \bra{x'} \mathcal{E}_{\rm eq}(\rho_{\rm in}) \ket{x'} dx' \\
=& \int_{x\in  \Delta}  \bra{r^{-1}x} \mathcal{E}_{\rm eq}(\rho_{\rm in}) \ket{r^{-1}x} r^{-1}dx \\
=& \int_{x\in  \Delta}  \frac{1}{ \sqrt{2\pi\frac{r^2}{4} } }\exp( -\frac{ (x-r\sqrt{\eta_e}x_0)^2 }{2\times \frac{r^2}{4}} ) dx.
\end{align}
Therefore, the condition for two models to be equivalent is

\begin{align}
r^2 &= 1+2\bar{n}(1-\eta_d) \label{homo_v},\\
\eta_e &=r^{-2}  \eta_d  . \label{homo_res}
\end{align}

\section{heterodyne measurement}\label{hetero}
The heterodyne measurement outputs a complex number $\omega\in \mathbb{C}$.
The probability density of an outcome $\omega$ of an ideal heterodyne measurement with an input state $\rho$ is given by $\frac{\braket{\omega|\rho|\omega}}{\pi}$. 
The probability distribution obtained by the trusted device model ${\rm Pr}\left( \omega\in \Delta \right)$ is written as

\begin{widetext}
\begin{align}
{\rm Tr}(\rho_{\rm in} \Pi_{\rm noisy}(\Delta))  & = {\rm Tr}( \mathcal{E}(\rho_{\rm in}) \Pi(\Delta)) \\
&=\int_{\omega\in\Delta}  \bra{\omega} \mathcal{E}(\rho_{\rm in}) \ket{\omega} \frac{d^2\omega}{\pi}\\
&= \int_{\omega\in\Delta} \frac{1}{2\pi\sigma^2_{\rm het}}\exp (-\frac{  (\omega_R-\sqrt{\eta_d}x_0)^2  + (\omega_I-\sqrt{\eta_d}p_0)^2 }{2 \sigma^2_{\rm het} }) d^2\omega
\end{align}
\end{widetext}
where we defined $\omega_R = {\rm Re}(\omega)$, $\omega_I = {\rm Im}(\omega)$, $p_0 = {\rm Im}(\alpha)$, and
\begin{align}
\sigma^2_{\rm het} & = \frac{1+ \bar{n} (1-\eta_d)}{2} .
\end{align}
Then, in the equivalent model, the probability distribution is given by
\begin{align}
&{\rm Tr}(\rho_{\rm in} \Pi_{\rm eq}(\Delta))  \\
=& {\rm Tr}( \mathcal{E}_{\rm eq}(\rho_{\rm in}) \Pi(r^{-1}(\Delta))) \\
=&\int_{\omega' \in r^{-1} (\Delta)}  \bra{\omega' } \mathcal{E}_{\rm eq}(\rho_{\rm in}) \ket{\omega' } \frac{d^2\omega' }{\pi} \\
=& \int_{\omega\in  \Delta}  \bra{r^{-1}\omega} \mathcal{E}_{\rm eq}(\rho_{\rm in}) \ket{r^{-1}\omega} \frac{d^2\omega}{r^2\pi} \\
=& \int_{\omega\in  \Delta}  \frac{1}{ 2\pi\frac{r^2}{2} }\exp( -\frac{ (x-r\sqrt{\eta_e}x_0)^2 }{2\times \frac{r^2}{2}} ) d^2\omega 
\end{align}

Therefore, the condition for two models to be equivalent is
\begin{align}
 r^2 &= 1+\bar{n}(1-\eta_d) \label{hetero_v}\\
\eta_e &= r^{-2}  \eta_d .  \label{hetero_res}
\end{align}


\bibliography{trust}


\end{document}